\documentclass[aps,pre,12pt,showpacs,%
superscriptaddress]{revtex4-1}
\usepackage{graphicx}
\usepackage[colorlinks=true,linkcolor=blue,%
pagecolor=blue,citecolor=blue,%
urlcolor=blue,anchorcolor=black,%
bookmarksnumbered=true,%
bookmarksopen=true]{hyperref}
\renewcommand{\frac}[2]{\displaystyle{#1 \over #2}}

\begin{document}
\title{Density distribution of a dust cloud in 
three-dimensional complex plasmas}
\author{V.~N.~Naumkin} \email{naumkin@ihed.ras.ru}
\author{D.~I.~Zhukhovitskii}
\author{V.~I.~Molotkov}
\author{A.~M.~Lipaev}
\author{V.~E.~Fortov}
\affiliation{Joint Institute of High Temperatures, Russian 
Academy of Sciences, Izhorskaya 13, Bd.~2, 125412 
Moscow, Russia}
\author{H.~M.~Thomas}
\author{P.~Huber}
\affiliation{Research Group Complex Plasma, DLR, 
Oberpfaffenhofen, 82234 Wessling, Germany}
\author{G.~E.~Morfill}
\affiliation{Max Planck Institute for extraterrestrial Physics, D-85748 
Garching, Germany}
\date{\today}
\begin{abstract}
We propose a novel method of determination of the dust 
particle spatial distribution in dust clouds that form in 
three-dimensional (3D) complex plasmas under microgravity 
conditions. The method utilizes the data obtained during the 
3D scanning of a cloud and provides a reasonably good 
accuracy. Based on this method, we investigate the particle 
density in a dust cloud realized in gas discharge 
plasma in the PK-3 Plus setup onboard the International Space 
Station. We find that the treated dust clouds are both 
anisotropic and inhomogeneous. One can isolate two regimes, 
in which a stationary dust cloud can be observed. At low 
pressures, the particle density decreases 
monotonically with the increase of the distance from the 
discharge center; at higher pressures, the density distribution 
has a shallow minimum. Regardless of the regime, we detect 
a cusp of the distribution at the void boundary and a slowly 
varying density at larger distances (in the foot region). A 
theoretical interpretation of obtained results is developed that 
leads to reasonable estimates of the densities for both the cusp 
and foot. The modified ionization equation of state, which 
allows for violation of the local quasineutrality in the cusp 
region, predicts the spatial distributions of ion and electron 
densities to be measured in future experiments.
\end{abstract}
\maketitle
\section{\label{s1}INTRODUCTION}
Complex plasmas are plasmas containing small solid 
particles, typically in the micrometer range, the so-called 
microparticles. These are dusty plasmas, which are 
specially prepared to study fundamental processes in the 
strong coupling regime on the most fundamental (kinetic) 
level through the observation of individual microparticles 
and their interactions~\cite{1,2,3,5,6,9}. Many interesting 
phenomena can be studied starting from small 
two-dimensional (2D) and three-dimensional (3D) 
clusters~\cite{1} to larger 2D and 3D systems where 
collective effects play a dominant role~\cite{6}.

Under laboratory conditions, the microparticles are heavily 
affected by the force of gravity. Gravity leads to the 
sedimentation of the particles and can be balanced either by 
a strong electric field in the sheath of a discharge or by the 
thermophoretic force due to a constant temperature gradient 
over the microparticle cloud. Additionally, there exist 
weaker forces like the neutral and ion drag forces and the 
interparticle interaction, which nevertheless can play a very 
important role in the total force balance.

Under microgravity conditions, e.g. onboard the 
International Space Station (ISS), gravity is negligible. 
Therefore, the particles are pushed out of the strong electric 
field region close to the electrodes due to their negative 
charge. They can form large and more or less homogenous 
particle clouds in the bulk of the 
discharge~\cite{10,15,16,17,18,019,19}. 
Under microgravity, weaker 
forces become important and define the motion and 
structure formation in complex plasmas.

The experiments performed under microgravity conditions 
allowed the researchers to obtain large 3D dusty plasma 
formations consisting of more than million dust 
particles~\cite{43}. However, these formations are not fully 
isotropic. This is a consequence of the plasma anisotropy 
along the chamber axis connected with the configuration of 
electrodes. Hence, a uniform distribution of dust particles in 
a dusty formation (dust cloud) can hardly be constructed.

The number density of dust particles is commonly 
estimated based on the interparticle distance in 
experimental 2D frames (see, e.g., \cite{11,19,43}). 
However, for lack of information on the type of a 
crystalline lattice of the dust crystal (in a strongly coupled 
particle system) and on the angle between the lattice axes 
and the beam of a laser used for optical diagnostics, such 
estimation is very crude so that it can define the number 
density only on the order of magnitude. 
In~\cite{17}, a standard tomography approach is applied for 
the determination of 3D particle coordinates in the dust 
cloud but this approach is not discussed. 
The methods of 3D structure determination
in complex plasmas were discussed in \cite{Klumov2010},
in particular, the bond order parameter method
\cite{Steinhardt1981} used to define the phase state
of a system under investigation \cite{17,43}.

The objective of this work is development of an accurate, 
reliable, and efficient method of determination of the spatial 
dust particle density distribution in complex 
plasmas and investigation of this distribution in dust clouds 
under different discharge conditions. We use the sequences 
of video frames taken in experiments with the 3D depth 
scans. A specially designed procedure allows one to restore 
the 3D particle coordinates defined as the coordinates of the 
center of mass of a cluster formed by individual dots in 2D 
frames. Then we find the local particle density 
using the Voronoi tessellation~\cite{Aurenhammer1991}. We find that two regions can 
be distinguished in each dust cloud, namely, the region of 
enhanced density adjacent to the space free from 
the particles (void), which we term the cusp, and the region 
of moderate variation of the density or the foot. We 
also find that the monotonic decrease of the density 
with the distance from the discharge center at the foot is 
replaced by an almost flat distribution with a shallow 
minimum in the foot center as the carrier gas pressure is 
increased. This is indicative of a change in the discharge 
structure.

For a foot, the theoretical interpretation of obtained results 
is provided by the ionization equation of state (IEOS) 
\cite{22,64}. However, this IEOS proves to be invalid for a 
cusp because the local quasineutrality condition implied in 
\cite{22,64} is violated in this region. To extend the IEOS 
to a cusp we replace this condition by constancy of the dust 
acoustic waves velocity across the dust cloud proved in 
\cite{64}. Thus, we arrive at the modified IEOS valid for 
both regions. Given the particle density, this 
equation allows one to predict the ion and electron  
densities, which can be measured in similar experiments 
with complex plasmas. Using the ``dust invariant''~\cite{64} 
we can provide a theoretical estimate based solely 
on the particle diameter and the electron temperature (at 
least, an order-of-magnitude one) for the particle  
density not only in the foot but also in the cusp region.

The paper is organized as follows. In Sec.~\ref{s2}, we 
discuss the design of experimental setup and the procedure 
of optical diagnostics of the dust cloud. In Sec.~\ref{s4}, 
we illustrate and analyze the results of particle  
density determination. The modified IEOS is derived in 
Sec.~\ref{s5}, the calculation results are compared with 
experimental data in Sec.~\ref{s6}. The results of this study 
are summarized in Sec.~\ref{s7}. In Appendices \ref{s8} 
and \ref{s9}, the method of experimental data processing is 
discussed in detail.

\section{\label{s2}EXPERIMENTAL SETUP AND 
PROCEDURE}
Experiments were performed in the PK-3 Plus laboratory 
onboard the ISS~\cite{18, Khrapak2016}. The heart of this 
laboratory is a parallel-plate radio-frequency (rf) discharge 
operating at a frequency of 13.56\,MHz, sketched in 
Fig.~\ref{f1}. The electrodes are circular plates with a 
diameter 6 cm made of aluminium. The distance between 
the electrodes is 3\,cm. The electrodes are surrounded by a 
1.5\,cm-wide ground shield including three microparticle 
dispensers on each side. The dispensers are magnetically 
driven pistons filled with monodisperse particles of various 
size and material~\cite{18}.

The discharge can operate in argon, neon, or their mixture in a 
wide range of pressures, rf amplitudes, and rf powers. The 
working pressures are in the range between 5 and 255\,Pa. 
Complex plasmas are formed by injecting monodisperse 
micron-size particles into the discharge. All the particles 
reaching the confinement region are kept there and form a 
complex plasma cloud. To observe the particles in the 
cloud, the optical detection system consisting of a laser 
illumination system and three video cameras (there is also a 
fourth camera of the same type which is used to observe 
glow characteristics of the plasma) is used. Two diode 
lasers ($\lambda = 686\,\mu$m) are collimated by a system 
of several lenses producing a laser sheet perpendicular to 
the electrodes with different opening angles and focal 
points. Three progressive CCD cameras detect the reflected 
light at $90^{\circ}$. An overview camera has a field of 
view (FoV) of about $60 \times 43$\,mm\textsuperscript{2} 
and observes the entire field between the electrodes. A 
second camera has a FoV of about $36 \times 
26$\,mm\textsuperscript{2} and observes the left part of the 
interelectrode space (about half of the entire system). The 
third camera is the high resolution camera with a FoV of 
about $8 \times 6$\,mm\textsuperscript{2}. It can be moved 
along the central axis in the vertical direction. The cameras 
and lasers are mounted on a horizontal translation stage 
allowing a depth scan through, and, therefore, 3D 
observation of complex plasma clouds. Further details on 
the PK-3 Plus project can be found in a comprehensive 
review~\cite{18}.

The experiments described here are carried out in argon at 
different pressures $p_{\mathrm{Ar}}$ and effective rf voltages 
$U_{eff}$ between electrodes (see Table~\ref{t:exp_par}).

We use melamine-formaldehyde particles. They were 
spheres with diameters presented in Table~\ref{t:exp_par}. 
The experimental procedure is as follows: when the 
particles form a stationary cloud in the bulk plasma, 
scanning was implemented by simultaneously moving laser 
and cameras in the direction perpendicular to the field of 
view with the velocity~0.6\,mm/s. Each scan 
takes~$\sim$\,8\,s, resulting in the scanning depth 
of~$\sim$\,4.8\,mm. The positions of dust particles were 
obtained on the video images of high resolution camera. 
The high resolution camera observes a region~$8 \times 
6$\,mm\textsuperscript{2} slightly above the discharge 
center and give the possibility to recognize separate dust 
particles. The processing procedure to find 3D particle 
positions will be presented in Appendix~\ref{s6}. 

\section{\label{s4} ANALYSIS OF OBTAINED DATA}
We use the experiment \# 2 (see Table~\ref{t:exp_par}) to 
obtain a representative spatial distribution of the dust 
particles density. Fig.~\ref{f2} presents video images of the 
dust cloud recorded by (a) the quadrant camera and (b) the 
high resolution camera.

Fig.~\ref{f3} demonstrates the 3D positions of dust 
particles according to the coordinates obtained using the 
method described in Appendix~\ref{s8}. The dust particles 
local density (in cm\textsuperscript{-3}) is indicated by 
color.

Using the experimental data presenting in Fig.~\ref{f3} we 
can determine positions of particles situated in the space 
between the parallel planes $z=0$ and $z=2$\,mm. As a 
result, we obtain the dependence of the dust local density 
$n_d$ on the coordinates $x$ and $y$ (Fig.~\ref{f3}). One can 
clearly see the chains of particles that correspond to the 
layers in the upper part of Fig.~\ref{f4}. Formation of the 
layers was noted in Ref.~\cite{Nefedov2003}. These layers 
lie in the plane $XZ$ parallel to the electrode plane. 
Apparently, the $Y$-axis perpendicular to these planes 
corresponds to a direction of the discharge electrical field. 
The orientation of observed layers testifies that the $Y$-
axis coincides with one of the dusty plasma crystallization 
axes. Note that the observed layers are formed near the 
electrode.

Figs.~\ref{f11}--\ref{f14} show the dependences of $n_d$
on the coordinate $y$. In this 
case, we take into account solely the particles situated at the 
distance from the $Y$-axis less than $r_s$. This distance 
restricts the number of particles involved in the density 
determination. For each particle diameter, $r_s$ 
corresponds to an optimum of sufficiently great number of 
particles and, at the same time, sufficiently small variance 
of the resulting particle density. For the selected values of 
$r_s$, this variance was always greater than the change of 
$n_d$ due to the cloud inhomogeneity inside 
the sampling volume in the directions perpendicular to the 
$Y$-axis. Note that in Figs.~\ref{f11}--\ref{f14}, 
reading of the $Y$-axis is directed from the plasma camera 
center so that on the left-hand side of each graph, one can 
see a region free from the dust particles termed the void.

A theoretical interpretation of the obtained data will be 
considered in the following section. This interpretation 
allows one to find a relation between the dust density 
obtained in the experiments with the densities of ions and 
electrons in plasma as well as to evaluate the densities of 
dust particles in different regions of the dust cloud.

\section{\label{s5}MODIFIED IEOS FOR THE 
THREE-DIMENSIONAL DUST CLOUD}

A dust particle in the cloud is subject to the following 
forces: the electric driving force arising from the ambipolar 
diffusion, the ion drag force due to the momentum transfer 
from streaming ions to the dust particles, the neutral drag 
force resulting from the collisions between moving particles 
and the gas molecules, and the force 
proportional to the gradient of the pressure of particle 
subsystem $p$. If we adopt the fluid approach developed in 
Ref.~\cite{64} then the fields of particle velocity 
${\bf{v}}(t,\,{\bf{r}})$
 and cloud density $\rho ({\bf{r}}) = Mn_d ({\bf{r}})$
 are solutions of the Euler equation
\begin{equation}
\frac{{\partial {\bf{v}}}}{{\partial t}} + 
({\bf{v}}\cdot\nabla ){\bf{v}} + \nu {\bf{v}} = 
\frac{1}{\rho }({\bf{f}}_e + {\bf{f}}_{id} - \nabla p) 
\label{e100}
\end{equation}
and the continuity equation
\begin{equation}
\frac{{\partial \rho }}{{\partial t}} + \nabla \cdot (\rho 
{\bf{v}}) = 0. \label{e200}
\end{equation}
Here, $\nu = (8\sqrt {2\pi } /3)\delta m_n n_n v_{T_n } a^2 
/M$
 is the friction coefficient defining the neutral drag 
\cite{33,6}, $\delta \simeq 1.4$
 is the accommodation coefficient corresponding to the 
diffuse scattering of ions against the particle surface, 
$m_n$ is the mass of a gas molecule, $n_n$ and $v_{T_n } 
= (T_n /m_n )^{1/2}$ are the number density and thermal 
velocity of the gas molecules, respectively, $T_n = 
300\;{\mbox{K}}$
 is the temperature of a gas, $a$
 is the particle radius, $M = (4\pi /3)\rho _d a^3$ is its 
mass, $\rho _d$ is the particle material density;
\begin{equation}
{\bf{f}}_e = - Zen_d {\bf{E}} = - \frac{{aT_e }}{e}\Phi 
n_d {\bf{E}} \label{e300}
\end{equation}
is the electric field driving force acting on unit volume 
where $Z$
 is the dust particle charge in units of the electron charge, 
$e$
 is the elementary electric charge, ${\bf{E}} = (T_e 
/e)\nabla \ln n_e$ is the electric field strength, $T_e$ is the 
electron temperature, $n_e$ is the electron number density, 
and $\Phi = - Ze^2 /aT_e$ is the dimensionless potential of 
a dust particle;
\begin{equation}
{\bf{f}}_{id} = \frac{3}{8}\left( {\frac{{4\pi }}{3}} 
\right)^{1/3} n_d^{1/3} n_i \lambda e{\bf{E}} 
\label{e400}
\end{equation}
is the ion drag force acting on unit volume where $n_i$ is 
the ion number density and $\lambda$ is the ion mean free 
path with respect to the collisions against gas atoms; and
\begin{equation}
p = \frac{1}{{8\pi }}\left( {\frac{{aT_e }}{{e\lambda ^2 
}}} \right)^2 p^* ,\quad p^* = \Phi ^2 n^{*4/3} 
\label{e500}
\end{equation}
is the dust pressure \cite{22}, where $n^* = (4\pi 
/3)\lambda ^3 n_d$ (in what follows, we will mark 
dimensionless quantities with an asterisk).

The nature of the ion drag force is rather complicated (e.g., 
\cite{65}) but the expression (\ref{e400}) we use is very 
simple. This expression is valid for a dense cloud where the 
Coulomb potentials of neighboring particles overlap. It 
estimates the momentum transfer cross section as the cross 
section of a sphere with the radius equal to the 
characteristic screening length in the Wigner--Seitz cell, 
which is $ \simeq 0.45(3/4\pi n_d )^{1/3} $. Although it 
does not distinguish between the collection and orbital parts 
of cross section, does not take into account the dependence 
of the cross section on the energy of an incident ion, effects 
of the Debye screening, the particle ordering, the ion wake 
formation, etc, it does depend explicitly on the local particle 
number density. It is this peculiarity of (\ref{e400}) that 
makes it possible to account for the existence of a 
stationary 3D particle cloud in a gas discharge. Note that 
the use of cross section for an isolated particle \cite{65}, 
which is obviously larger than (\ref{e400}), cannot explain 
the cloud formation because for all particles, the force 
balance is reached only at some surface. It is worth 
mentioning that formula (\ref{e400}) agrees well with 
experiment. The ion drag force acting on an individual 
particle was directly measured in Ref.~\cite{66}. For the 
particles of the radius $1.25 \times 10^{ - 4} 
\;{\mbox{cm}}$
 used in this experiment, our approach is valid, and the 
force per one particle ${\bf{f}}_{id} /n_d$ calculated from 
(\ref{e400}) amounts to $5.4 \times 10^{ - 9} 
\;{\mbox{dyn}}$. This lies within the experimental error 
estimated in \cite{66} and is less than the force calculated 
using the cross section for an isolated dust particle.

For a stationary cloud, the neutral drag vanishes. The 
particle pressure force can be neglected in most cases 
provided that the condition $a\Phi T_e^2 /Mc^2 e^2 \gg 1$
 is satisfied \cite{64}. Here, $c$
 is the velocity of DAWs propagation (sound velocity). A 
stationary state of the dust cloud implies the zero total force 
acting on each dust particle. Therefore, for a stationary 
cloud, the electric driving force is compensated by the ion 
drag, ${\bf{f}}_e + {\bf{f}}_{id} = 0$. This force balance 
equation is reduced to \cite{22,64}
\begin{equation}
\left( {\frac{{9\pi }}{{128}}} \right)^{1/3} \frac{{n_i 
\lambda }}{{n_d^{2/3} }} = \frac{{aT_e }}{{e^2 }}\Phi , 
\label{e8}
\end{equation}
It is completed by the equation defining the particle 
potential
\begin{equation}
n_e = n_i \theta \Phi e^\Phi , \label{e9}
\end{equation}
where $\theta = \sqrt {T_e m_e /T_i m_i } $, $T_i$ and 
$m_i$ are the ion temperature and mass, respectively, and 
$m_e$ is the electron mass, and by the local quasineutrality 
condition
\begin{equation}
n_i = \frac{{aT_e }}{{e^2 }}\Phi n_d + n_e . \label{e600}
\end{equation}
Note that (\ref{e9}) implies the OML model \cite{54,55}, 
which does not include the ion collisions in the vicinity of 
the particles resulting in the particle charge reduction 
\cite{67}. However under the conditions of our experiment, 
this effect seems to be negligible. Thus for $2a = 6.8 \times 
10^{ - 4} \;{\mbox{cm}}$
 (Sec.~\ref{s6}), the charge decrease by 30\% leads to the 
decrease in the ion and electron number densities on the 
same order of magnitude, which is knowingly less than the 
accuracy of the theory. In addition, this effect is ignorable 
for the Havnes numbers greater than unity (see the 
discussion in \cite{64}), which is also typical for our 
experiment. 

The resulting IEOS for a dust cloud can be written in the 
form of a local relation between each pair of the following 
plasma ionization state parameters: the electron, ion, and 
particle number density, and the particle potential 
\cite{22,64}. For the variables $n_i$ and $\Phi $, the IEOS 
is
\begin{equation}
n_i^* = \frac{{128}}{{9\pi }}\Phi \left( {1 - \theta \Phi 
e^\Phi } \right)^2 , \label{e10}
\end{equation}
for $n_d$ and $\Phi $, 
\begin{equation}
n^* = \frac{{512}}{{27}}(1 - \theta \Phi e^\Phi )^3 , 
\label{e11}
\end{equation}
where $n_i^* = (e^2 \lambda ^3 /aT_e )n_i $. 
A simple relation between $n^* $, $n_i^* $, and $\Phi$ 
follows from (\ref{e10}) and (\ref{e11}),
\begin{equation}
n_i^* = \frac{2}{\pi }\Phi n^{*2/3} . \label{e111}
\end{equation}

The DAWs in a stationary dust cloud can be 
treated on the basis of Eqs.~(\ref{e100}) and (\ref{e200}) 
linearized with respect to small variations of ${\bf{v}}$
 and $\rho $:
\begin{equation}
\frac{{\partial ^2 \psi }}{{\partial t^2 }} + \nu 
\frac{{\partial \psi }}{{\partial t}} = c^2 \Delta \psi , 
\label{e700}
\end{equation}
where $\nabla \psi = {\bf{v}}$
 and $c^2 = dp/d\rho $. The latter derivative is calculated 
using Eqs.~(\ref{e8}), (\ref{e9}), and (\ref{e600}),
\begin{equation}
c = \frac{{aT_e c^* }}{{e\sqrt {6M\lambda } }},\quad 
c^{*2} = \frac{4}{3}\Phi ^2 n^{*1/3} \left[ {1 - 
\frac{3}{2}\frac{{n^{*1/3} }}{{(\Phi + 1)\left( {8 - 
3n^{*1/3} } \right)}}} \right], \label{e800}
\end{equation}
where $n^{*1/3} = (8/3)(1 - \theta \Phi e^\Phi )$. The 
sound velocity (\ref{e800}) proves to be almost 
independent on the coordinate. Hence, a good estimate is 
provided at some point inside a dust cloud, e.g., at the point 
where the condition \begin{equation}
dn_e /dn_i = 0
\end{equation} is satisfied. At this point, the particle 
potential $\Phi _s$ is defined by the equation \cite{64}
\begin{equation}
\frac{{2\theta e^{\Phi _s } (\Phi _s + 1)}}{{1 - \theta \Phi 
_s e^{\Phi _s } }} = 1 + \frac{2}{{\Phi _s }} \label{e14}
\end{equation}
and then
\begin{equation}
c^{*2} = \frac{{512}}{{27}}\frac{{\Phi _s^2 (\Phi _s + 
1)}}{{(3\Phi _s + 4)(\Phi _s + 2)}}. \label{e13}
\end{equation}

Equations (\ref{e10}) and (\ref{e11}) imply the local 
quasineutrality of plasma, which takes place provided that 
the screening length is much smaller than the length scale 
of plasma state parameters variation. In our experiments, 
this is true for most part of the cloud, where the change of 
$n_d$ is insignificant. In what follows, we will term this 
part of the cloud the foot. At the same time, as the void 
boundary is approached, $n_d$ increases sharply in a 
relatively narrow region (cusp), whose width is on the same 
order of magnitude as the largest screening length of this 
system, the electron Debye length (see Sec.~\ref{s5}). 
Therefore, Eqs.~(\ref{e10}) and (\ref{e11}) are not 
applicable in this region of the cloud due to the 
quasineutrality violation. If we replace the local 
quasineutrality condition in this region by the Poisson 
equation, the problem would be reduced to solution of a set 
of differential equation, and a local solution would no 
longer be possible. However, we can replace the 
quasineutrality condition by another condition to find a 
local form of an extended IEOS. According to the results of 
experimental determination of the sound velocity in argon 
\cite{19} and neon \cite{35}, the latter is independent of 
the coordinate inside the dust cloud including the void 
boundary. Also, no variation of the measured sound 
velocity within the experimental accuracy was revealed in 
all available experimental studies. Note that IEOS 
(\ref{e10}) and (\ref{e11}) lead to a sharp vanishing of the 
sound velocity in a close vicinity of the void, which is an 
obvious consequence of the quasineutrality violation not 
taken into account in Ref.~\cite{64}.

The quasineutrality condition that preserves the locality of 
IEOS can be replaced by the requirement of constancy of 
the sound velocity $c$, 
\begin{equation}
\frac{{dp^* }}{{d\rho ^* }} = c^{*2} = 
{\mbox{const}}{\mbox{.}} \label{e12}
\end{equation}
The solution of Eq.~(\ref{e12}) has the form
\begin{equation}
\Phi = c^* \frac{{\sqrt {n^* - n_0^* } }}{{n^{*2/3} }}, 
\label{e15}
\end{equation}
where $n_0^*$ is the integration constant to be determined. 
We substitute (\ref{e15}) into (\ref{e8}) to derive the 
relation between $n_i^*$ and $n^* $,
\begin{equation}
n_i^* = \frac{2}{\pi }c^* \sqrt {n^* - n_0^* } . \label{e16}
\end{equation}

Obviously, for the ion and particle number density 
corresponding to the transition from the cusp to the foot, the 
right-hand side of relations (\ref{e111}), where $\Phi$ is a 
solution of equation (\ref{e11}), and (\ref{e16}) must be 
equal. If we denote the corresponding transition 
dimensionless particle number density and potential by 
$n_f^*$ and $\Phi _f $, respectively, then it follows from 
(\ref{e111}) and (\ref{e16}) that
\begin{equation}
n_0^* = n_f^* \left( {1 - \frac{{\Phi _f^2 n_f^{*1/3} 
}}{{c^{*2} }}} \right). \label{e17}
\end{equation}
We use the experimentally determined $n^*$ to estimate 
the ion number density in the dust cloud on the basis of the 
IEOS (\ref{e16}) and (\ref{e17}). Given the ion and 
particle number density, the electron number density is 
calculated using Eqs.~(\ref{e9}), (\ref{e15}), and 
(\ref{e17}).

The transition particle number density $n_{df} = 3n_f^* 
/4\pi \lambda ^3$ can be estimated using the ``dust 
invariant'' \cite{22}
\begin{equation}
\kappa = \left( {\frac{3}{{4\pi }}} \right)^{2/3} 
\frac{1}{{aT_e n_{df}^{2/3} }}, \label{e18}
\end{equation}
which is not much different from $\kappa = 
0.209\;{\mbox{cm/eV}}$
 (if $a > 2 \times 10^{ - 4} \;{\mbox{cm}}$) for all 
available experimental data obtained by different authors 
(e.g., \cite{43,11,19,37}). It is reasonable 
to assume that for the maximum particle number density at 
the cusp top $n_{dc} $, the quantity
\begin{equation}
\kappa ' = \left( {\frac{3}{{4\pi }}} \right)^{2/3} 
\frac{1}{{aT_e n_{dc}^{2/3} }} \label{e19}
\end{equation}
is also almost independent of the experimental conditions.

\section{\label{s6}DISCUSSION}

Distributions of the particle density along the 
$Y$-axis passing through the discharge (void) center 
perpendicular to the electrodes are shown in 
Figs.~\ref{f11}--\ref{f14} for different particle diameter 
$2a$
 and argon pressure $p_{\mathrm{Ar}} $. It is clearly seen 
in the figures that the dust cloud is divided in two regions. 
In the void boundary, a fast decrease of the density 
with the increase of the coordinate $y$
 is observed, and a cusp is formed. Farther from the 
discharge center, the dependence $n_d (y)$
 is rather weak. In Sec.~\ref{s5}, this region was termed a 
foot.

In the foot region, the behavior of $n_d (y)$
 changes qualitatively as the argon pressure is increased 
from $10$
 to $20\;{\mbox{Pa}}$. In Figs.~\ref{f11} and \ref{f12}, 
$n_d (y)$
 decreases monotonically while in Figs.~\ref{f13} and 
\ref{f14}, $n_d (y)$
 has a wide shallow minimum approximately in the foot 
center. Apparently, this is related to the change in the 
discharge regime stimulated by presence of the dust 
particles. Regardless of the regime, the characteristic 
particle density in the foot region is in a 
satisfactory agreement with the estimates based on the 
``dust invariant'' $\kappa = 0.209\;{\mbox{cm/eV}}$
 (\ref{e18}) (large solid green circles in 
Figs.~\ref{f11}--\ref{f14}). If we set the ``cusp invariant'' 
$\kappa '$
 (\ref{e19}) to $0.129$, we obtain a reasonable estimates 
for the maximum particle density (in 
Figs.~\ref{f11}--\ref{f14}, such density is 
indicated by large solid magenta circles).

Note that for $2a = 6.8 \times 10^{ - 4} \;{\mbox{cm}}$
 (Fig.~\ref{f14}), the void boundary was not stationary but 
it was involved in the heartbeat instability~\cite{18}, which 
makes problematic the determination of particle coordinates 
due to elongation of the visible tracks of rapidly oscillating 
particles. Thus, the particle density determined in 
this region is questionable. Also, the theoretical estimations 
of Sec.~\ref{s5} cannot be applied for an unstable system. 
However, we take into account that the unstable region is 
very narrow and the most part of the cloud is steady state, 
and hence, we neglect the instability.

Based on the experimentally determined
particle density 
distributions, we can estimate the ion and electron  
densities. The experimental data on $n_d$ were fitted by 
different quadratic trinomials in the cusp and foot regions 
(see Figs.~\ref{f11}--\ref{f14}). Obtained dependences 
$n_d (y)$
 were then used for the calculation of $n_i (y)$
 (\ref{e16}) and $n_e (y)$
 (\ref{e9}) (Figs.~\ref{f15} and \ref{f16}). As is seen, the 
dependences $n_i (y)$
 and $n_e (y)$
 are sensitive to the argon pressure but not much sensitive 
to the particle diameter. Theoretical estimates presented in 
these figures point to the ion and electron  densities 
on the order $10^9 \;{\mbox{cm}}^{ - 3} $. For $n_i (y)$, 
the cusp and foot regions are clearly separated in 
Fig.~\ref{f15} due to an explicit dependence of $n_i$ on 
$n_d$ in relation (\ref{e16}). For $p_{\mathrm{Ar}} = 
20.5\;{\mbox{Pa}}$, the dependences have shallow 
minima stipulated by similar minima in the dependence 
$n_d (y)$. Apparently, the depth of these 
minima is significantly smaller than the accuracy of the 
theory. If we judge by the correspondence between the 
measured and estimated particle number density in 
\cite{22} then the accuracy within an order of magnitude 
can be expected. Therefore, the minima of $n_i (y)$
and $n_e (y)$ may not be real.

Figure \ref{f16} shows the same peculiarity for $n_e (y)$. 
The minimum of $n_e$ signifies the inversion of the 
electric field strength ${\bf{E}} = (T_e /e)\nabla \ln n_e$ 
and of the ion flux direction. At the same time, the curves 
$n_e (y)$
 are shifted down relative to $n_i (y)$
 due to the negative charge of dust particles. This effect is 
especially strong for the smallest particles $2a = 2.55 \times 
10^{ - 4} \;{\mbox{cm}}$
 ($n_i /n_e \sim 10$), which is a result of the increase of 
 density with the decrease of particle diameter (cf.\ 
(\ref{e18})). Note that for this case, in the cusp region, 
Eq.~(\ref{e111}) is inapplicable because $n^*$ exceeds the 
maximum particle density that can be reached in 
the locally quasineutral system~\cite{64}. This is not 
surprising because the electron Debye length $l_D = \sqrt 
{T_e /4\pi n_{df} e^2 } \simeq 0.1\;{\mbox{cm}}$
 is close to the cusp width, so that quasineutrality is 
significantly violated in the cusp region. In contrast, for 
larger particles, the calculation using formulas (\ref{e111}) 
and (\ref{e16}) leads to little different results even in the 
cusp region. Here, $l_D$ ranges from $2.8 \times 10^{ - 
5}$ to $5.5 \times 10^{ - 5} \;{\mbox{cm}}$, which is 
smaller than the cusp length, so that the quasineutrality 
approximation is acceptable.

Unfortunately, at present, no reliable data on the ion and 
electron  densities in discharge complex plasmas are 
available. Thus, the estimates made in this Section can 
serve a guide for further experiments. Although the IEOS is 
formulated in a convenient form of local relations, the 
accuracy of applied theory is limited by an order of 
magnitude (however, one can hope that the above-discussed 
qualitative peculiarities would hold). The accuracy of a 
theory can be increased by numerical solution of exact 
equations that are relevant for formation of a dust cloud in 
complex plasmas.

\section{\label{s7} CONCLUSION}

In this study, we have developed a novel method of 
investigation of the dust particle density 
distribution in complex plasmas. This method implies 
digital processing the video frames taken during 3D 
scanning of a dust cloud. The coordinates of an individual 
particle are assigned to those of the center-of-mass of a 
cluster comprised of the corresponding dots resolved on 
video frames. The local dust density at the point of 
particle location is defined as the inverse volume of the 
Voronoi cell for this particle. We have found that this 
method provides the most accurate information on the 
spatial particle density distribution. We performed 
the investigation of dust clouds in strongly coupled 
complex plasmas (plasma crystals) based on this method.

The plasma crystals proved to be essentially 
inhomogeneous and anisotropic formations. For all 
experimentally observed plasma crystals, a strong 
anisotropy is caused by the design of experimental setup. 
The axial symmetry of the camera results in formation of 
the principal axis of anisotropy ($Y$-axis) passing through 
the camera center perpendicular to the plane of electrodes. 
The proposed method makes it possible to resolve clearly the 
crystalline lattice planes of a dust crystal parallel to the 
electrodes. The lattice planes are perpendicular to the 
direction of the electric field of ambipolar diffusion. 
Analyses of the particle density distribution shows 
that its variation is much stronger in the direction of 
$Y$-axis than along other directions. We find that the 
density increases significantly at the void boundary 
forming a cusp of the distribution, while sufficiently far 
from the discharge center (in the foot region), the  
density changes moderately. 

Investigation of the dust cloud at different argon pressures 
and electrode voltage for three particle diameters revealed 
two typical modes of dusty plasma realization. Regardless 
of the particle diameter, the mode changes with argon 
pressure variation. At lower pressures, the foot particle 
 density decreases monotonically with the increase 
of the distance from the discharge center. In contrast, at 
higher argon pressures, the foot density is almost 
constant. In this case, the density distribution has a shallow 
minimum whose depth increases with the increase of the 
pressure. Apparently, there exists a threshold pressure, for 
which the structure of discharge complex plasma is 
rearranged.

We have modified the IEOS for complex plasmas with a 
similarity property to extend the theoretical interpretation of 
obtained density distributions to the cusp region 
where the assumption of local quasineutrality flaws. The 
modified IEOS that explores constancy of the dust acoustic 
waves velocity allows one to predict the profiles of electron 
and ion densities in complex plasmas. The 
characteristic cusp and foot particle densities are in 
a satisfactory agreement with the estimates based on the 
``dust invariants''.

The developed method can be used for various realizations 
of complex plasmas. Exhaustive information on the 
structure of a dust cloud could be gained in future 
experiments with the increased camera field of view and 
scanning depth including almost the whole dust cloud. 
Simultaneous measurements of the ion and electron 
densities would be highly important.

\begin{acknowledgments}
The support of the Russian Science Foundation for 
development of the diagnostic technique and processing the 
experimental data (Project No.~14-50-00124, V.\ N.\ N., 
V.\ I.\ M., and A.\ M.\ L.) and for development of the 
extended IEOS for the discharge dusty plasmas (Project 
No.~14-50-00124, D.\ I.\ Zh.) and of the DLR/BMWi for 
conduct of the experiments (Grants No.~50WM0203 and 
50WM1203, H.\ M.\ Th.\ and P.\ H.) are gratefully 
acknowledged.
\end{acknowledgments}

\appendix
\section{\label{s8} Method of the 3D coordinates of dust 
particles determination}
We used video images of the high resolution camera 
because other cameras do not make it possible to reliably 
resolve individual particles. The experimental data in 
digital form obtained during the dusty cloud scanning were 
stored on the hard drive and processed in the form of the 
3D matrix $M_{i,j,k}$ where the subscripts $i$ and $j$ 
correspond to the coordinates in the video frame plane (to 
the axes $Y$ and $X$, respectively) and the subscript $k$ 
corresponds to the frame number (or to the $Z$-axis). The 
elements of this matrix are the 3D pixels which are 
conventionally called voxels. The voxel size is determined 
by the size of the pixel in the frame and the value of the 
laser sheet displacement.

First, the 3D Gauss filter (see~\cite{Nixon2008}) with 
parameter $\sigma_G$ was applied to the matrix 
$M_{i,j,k}$. The objective of this procedure is to reduce 
the noise level. Then, the 2D matrix of minimums of the 
brightness along the $X$ and $Y$-axes was calculated and 
this matrix was subtracted from $M_{i,j,k}$ along the 
$Z$-axis. We perform this procedure to eliminate the 
constant component in all frames and to reduce the 
influence of the pixels on the matrix of video camera with 
the increased level of brightness.

The voxels with the brightness above threshold $I_t$ are 
gathered in clusters. By definition, the voxel belongs to 
given cluster if it has at least one neighbor comprising the 
same cluster corresponding to the subscripts $i$, $j$, and 
$k$ that differ from the subscripts of this voxel no more 
than by some $\xi$. The brightness threshold as well as the 
value of $\xi$ were chosen manually taking into account 
the best visual correspondence of the images of the particles 
in the frames and their centers determined on the basis of 
algorithm. Parameters of data processing are listed in 
Table~\ref{t:par_dat}.

Typically, the threshold was a little bit higher than the level 
of the background brightness and $\xi = 3$.

3D coordinates of the particle are determined as coordinates 
of ``the center of mass'' of the cluster corresponding to this 
particle with due regard for the brightness of the voxels as a 
weight factor. In this work, we form a block containing 3D 
coordinates of the dust particles in the region corresponding 
to the observation area of the high-resolution camera and to 
the scanning depth.

Given the frequency of video exposure (50\,frames/s) and 
the velocity of the platform movement (0.6\,mm/s) during 
the scanning, it is possible to calculate the laser sheet 
displacement between two successive frames (12\,$\mu$m). 
For the high resolution camera, the pixel size in the frame is 
equal to $11.3 \times 10.3\,\mu$m. For the sake of 
convenience of subsequent analysis of the results, we 
assume that the plasma camera center is the center of the 
coordinate system.

\section{\label{s9} Local density of the dust particles}
\begin{figure}
\includegraphics[width=18cm]{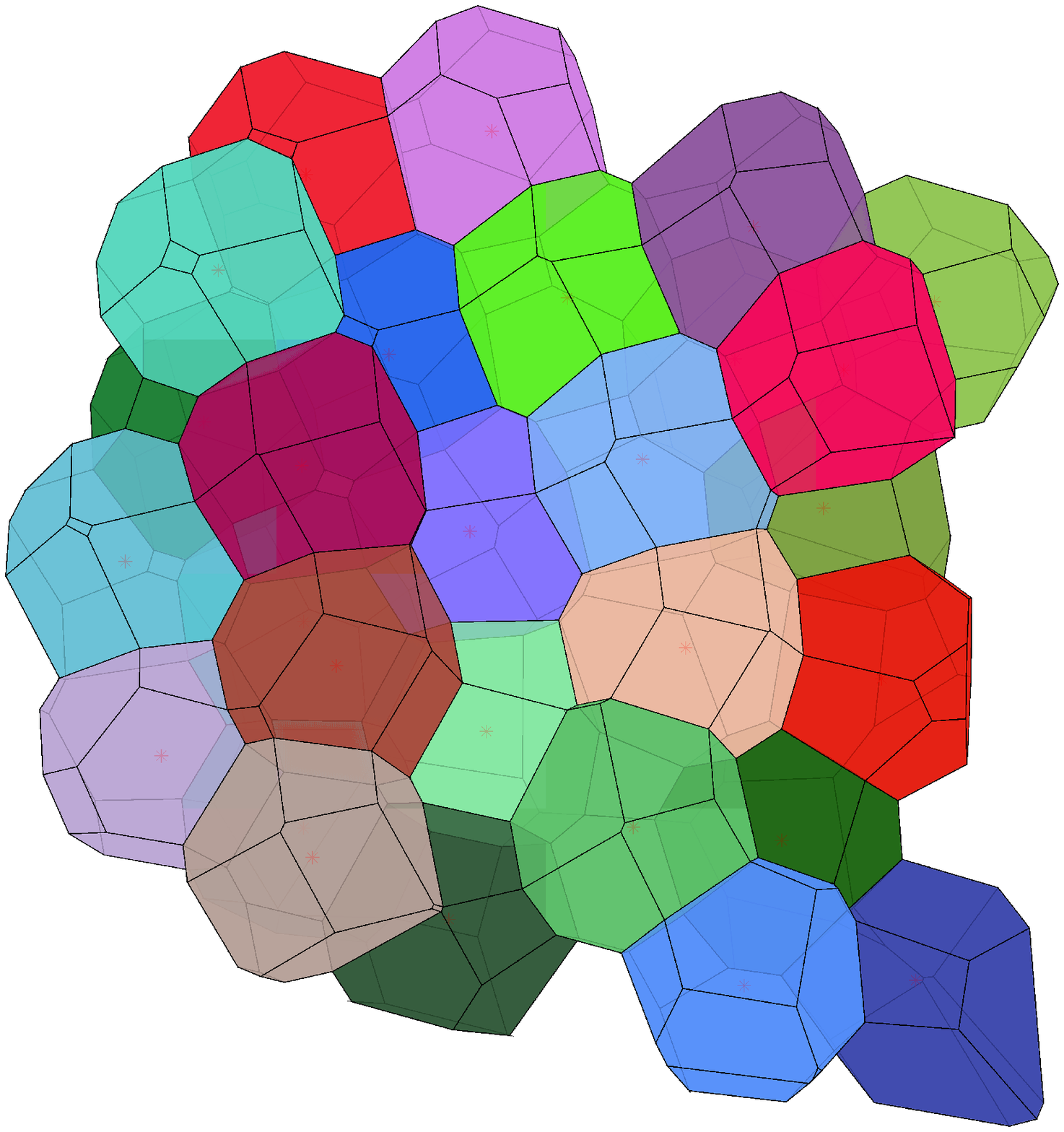}
\caption{\label{f_voronoi}}
\end{figure}

The objective of this paper is a determination of the dust 
particles density distribution in the region accessible for the 
high-resolution camera. Here, we perform data processing 
for the region of the dust cloud that is situated between the 
void boundary and the near electrode region. This region is 
characterized by the nonuniform distribution of the dust 
particles. To find the local density of dust particles we use 
3D coordinates of the dust particles determined on the basis 
of the procedure discussed in Appendix~\ref{s8}. We use 
these coordinates to construct the 3D Voronoi diagram, 
which is constructed of a set of the Voronoi cells~\cite{Aurenhammer1991}. 
Fig.~\ref{f_voronoi} illustrates a view of such set in space.

\begin{figure}
\includegraphics[width=18cm]{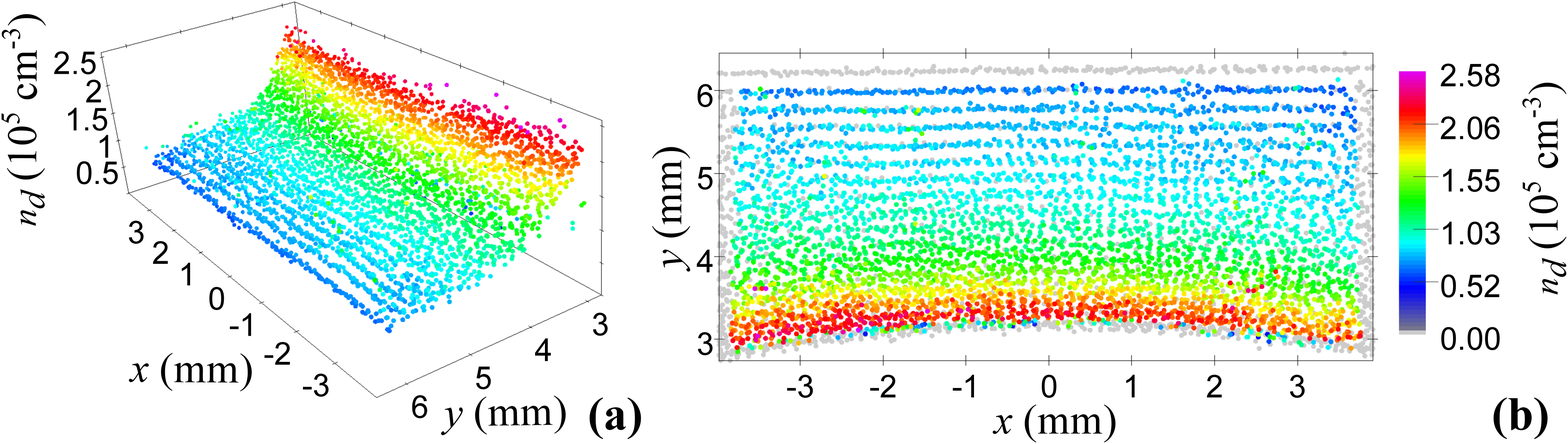}
\caption{\label{f4}}
\end{figure}

\begin{figure}
\includegraphics[width=18cm]{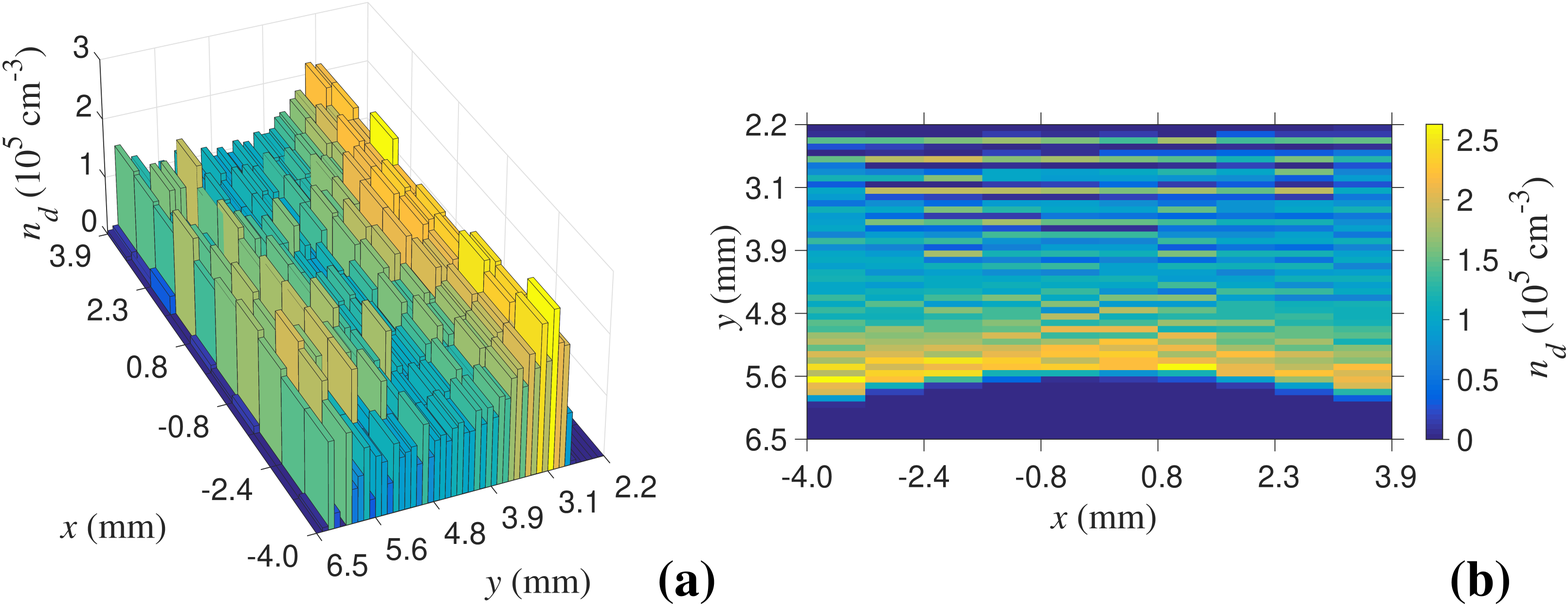}
\caption{\label{f_num}}
\end{figure}

We estimate the particle local density as the inverse volume 
of the 3D Voronoi cell. Such processing allows one to 
assign the volume of Voronoi cell~$V_i$ and the 
corresponding local density $n_i=1/V_i$ to each dust 
particle with the coordinates $x$, $y$, and $z$. This allows 
us to get more detailed information on the distribution of 
dust particles density in comparison with the commonly 
used number density $n=N/V$, where $N$ is the number of 
particles in the volume~$V$. To compare two methods of 
obtaining dust density distribution we present in 
Fig.~\ref{f_num} the distribution of dust number density 
for the experimental conditions corresponding to 
Fig.~\ref{f4} where the distribution of local particle 
density is shown. This example testifies that using the 
Voronoi tessellation makes it possible to obtain more 
detailed information concerning the particle distribution.

It should be noted that for the particles situated at the outer 
boundaries of the cloud, the Voronoi cells are not closed; 
therefore, it is impossible to determine the local density. 
Moreover, in calculation of the Voronoi cell volume, we 
have to exclude the cell tops that are beyond the limits of 
the region of the coordinate determination. Hence, we do 
not include in the analysis the highly deformed cells, which 
can be encountered near the cloud boundary.

\begin{figure}
\includegraphics[width=18cm]{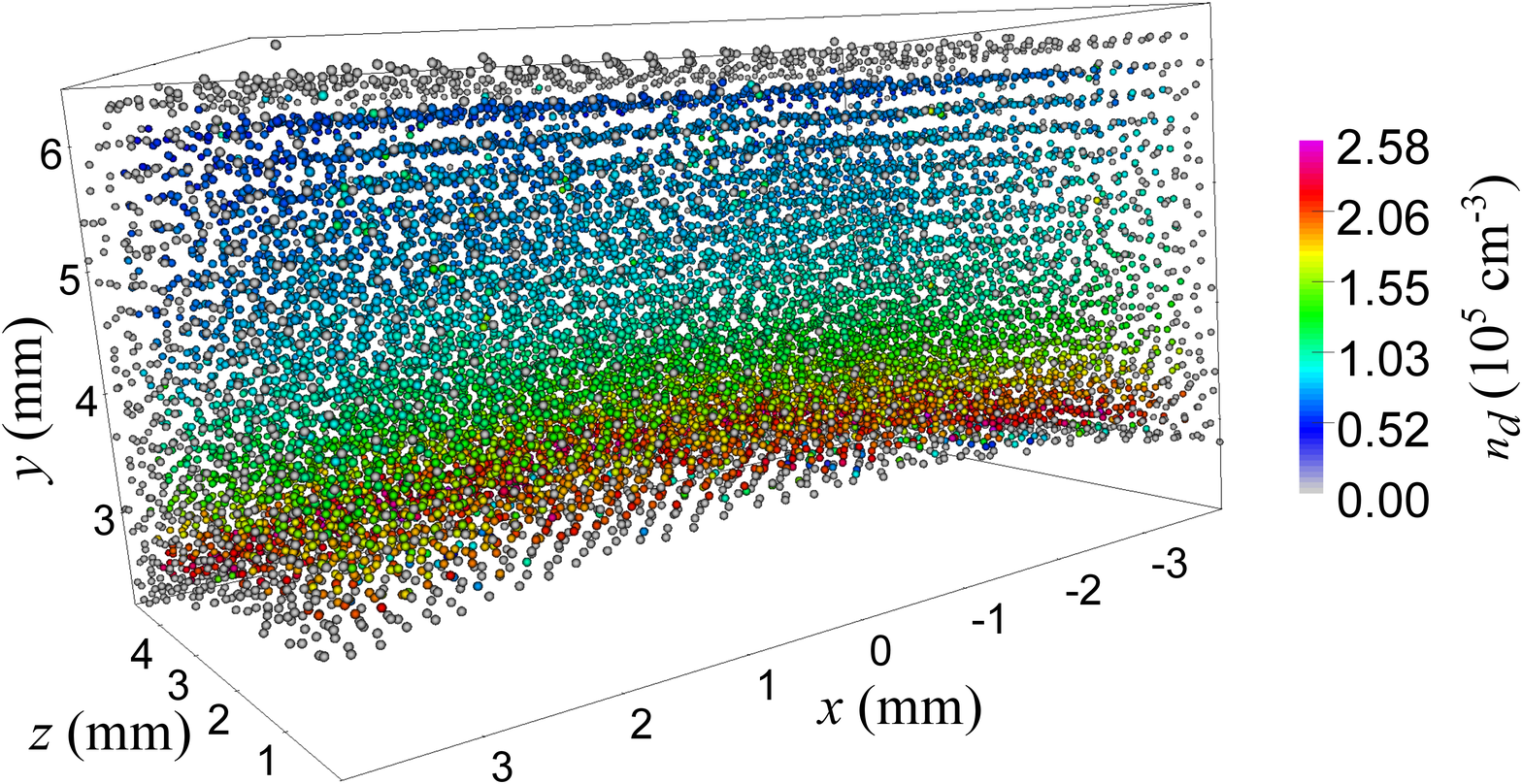}
\caption{\label{f3}}
\end{figure}

\begin{figure}
\includegraphics[width=18cm]{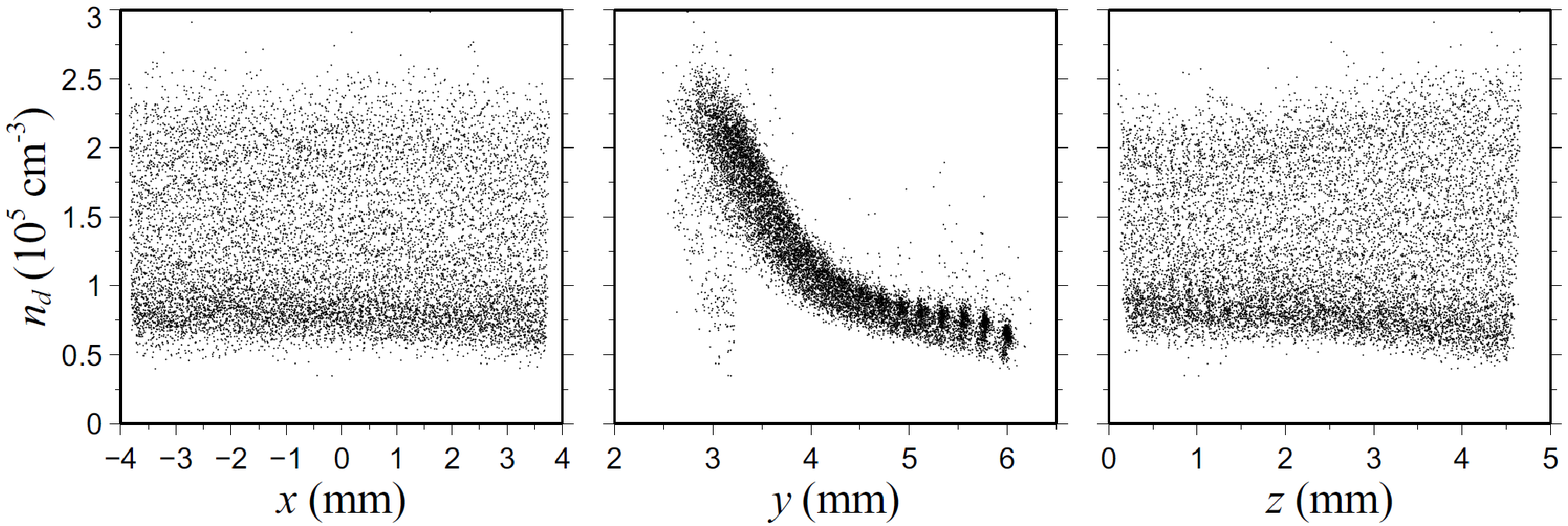}
\caption{\label{f_plots}}
\end{figure}

The determined dust particles local densities are shown vs.\ 
the particle coordinates $x$, $y$, and $z$ in the plots 
Fig.~\ref{f_plots}. The processing was made 
for the experimental conditions corresponding to 
Fig.~\ref{f3}. A correlation between the dust local  
density and the coordinate on $Y$-axis, which is the central 
axis of the plasma chamber, is clearly seen.

\newpage
\providecommand{\noopsort}[1]{}\providecommand{\singleletter}[1]{#1}%

\clearpage

\begin{table}
	\caption{\label{t:exp_par} Experiment parameters.}
	\begin{ruledtabular}
		\begin{tabular}{llll}
			Experiment \# & $d_p$, $\mu$m & $p_{\mathrm{Ar}}$, Pa & $U_{eff}$, V \\
			\hline
			1 & 2.55 & 10 & 14.5 \\ 
			2 & 3.42 & 11 & 14.6 \\ 
			3 & 3.42 & 20.5 & 13.2 \\ 
			4 & 6.8 & 20.5 & 14.2 \\
		\end{tabular}
	\end{ruledtabular}
\end{table}

\begin{table}
	\caption{\label{t:par_dat} Parameters of data analysis.}
	\begin{ruledtabular}
		\begin{tabular}{lllll}
			Experiment \#	& $\sigma_G$	& $I_t$	& $\xi$	& $r_s$, mm	\\
			\hline
			1	& 1	& 35	& 2	& 1	\\
			2	& 1	& 30	& 3	& 1.7	\\
			3	& 1	& 30	& 3	& 1.7	\\
			4	& 2	& 35	& 4	& 2	\\
		\end{tabular}
	\end{ruledtabular}
\end{table}

\newpage
\clearpage

\centerline{FIGURE CAPTIONS}
\vskip\baselineskip

\begin{figure}
	\includegraphics[width=18cm]{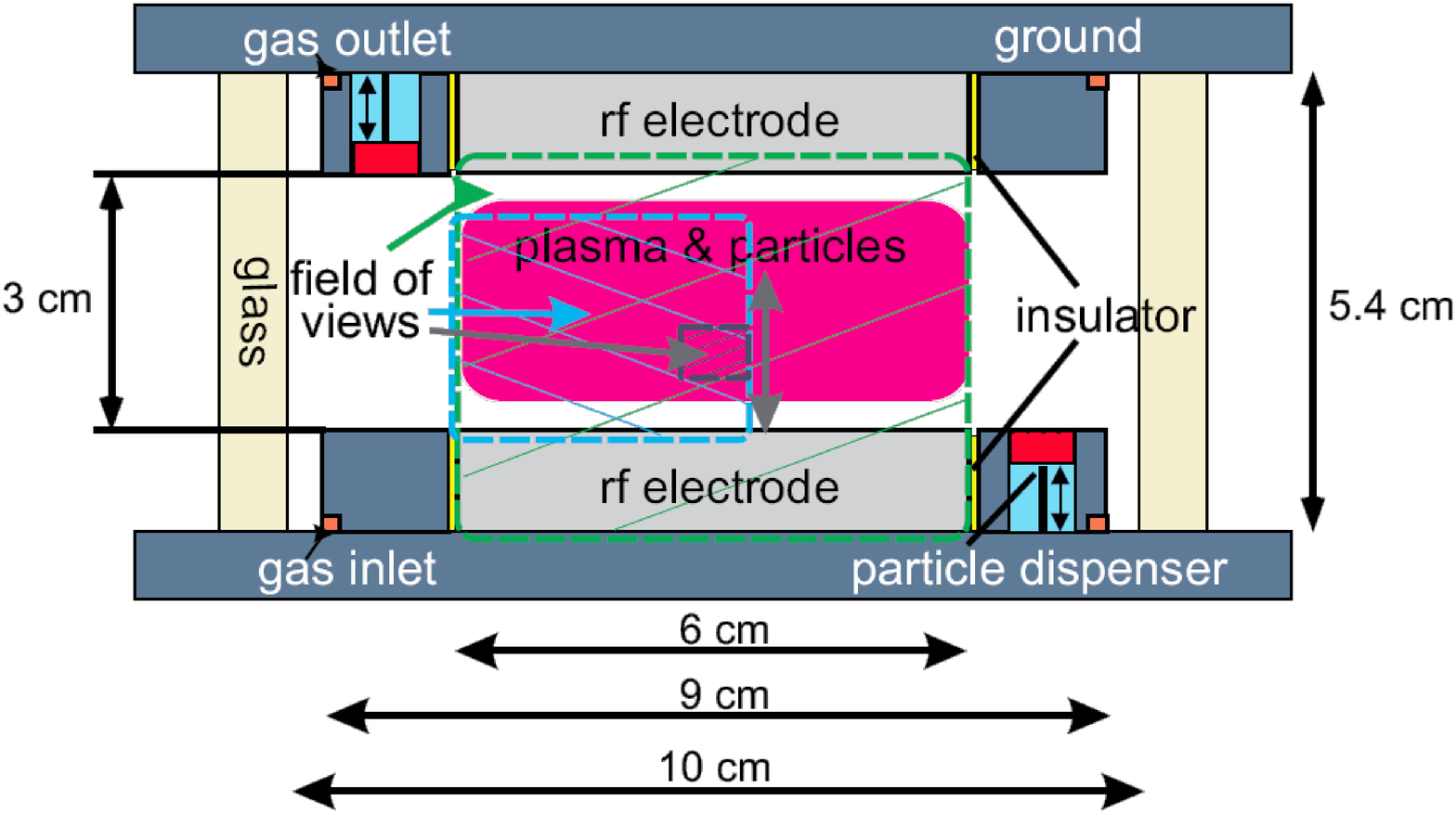}
	\caption{\label{f1}}
\end{figure}

Fig.~\ref{f1}: (Color online) Sketch of the PK-3 Plus 
plasma chamber~\cite{18}.

\begin{figure}
	\includegraphics[width=18cm]{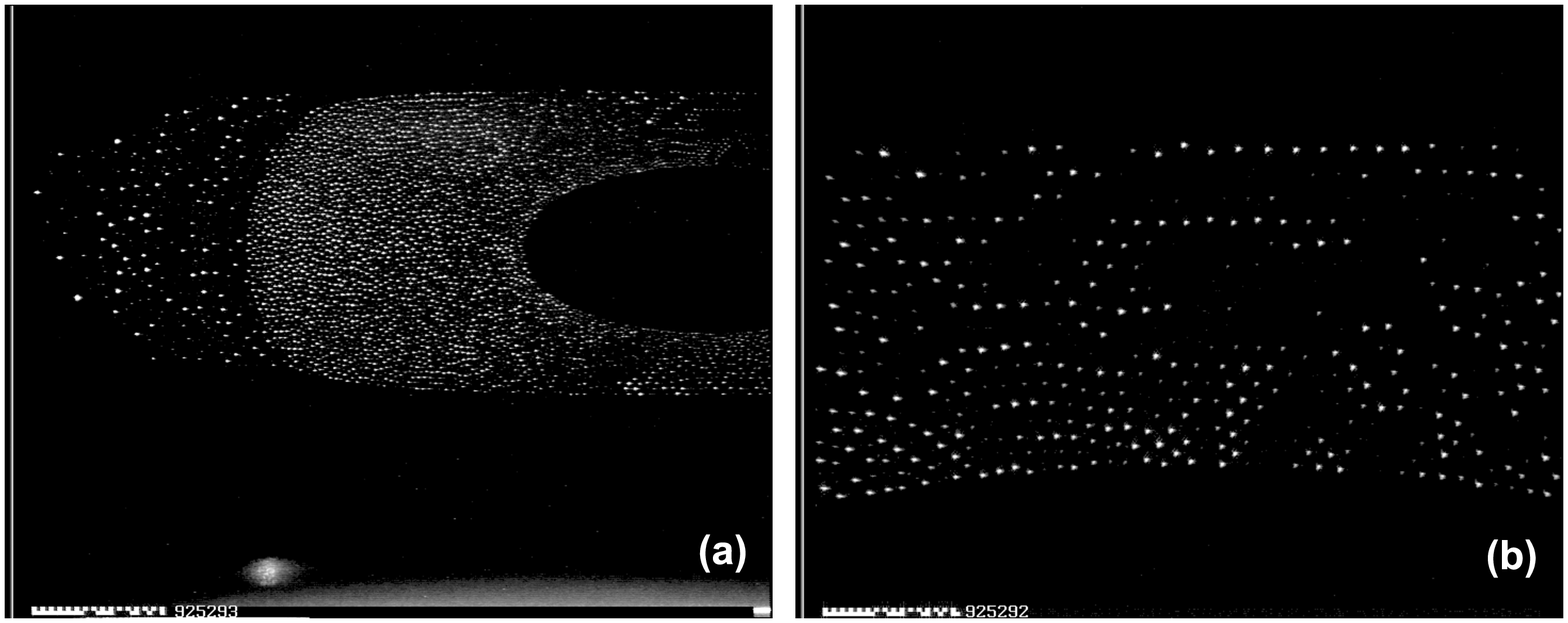}
	\caption{\label{f2}}
\end{figure}

Fig.~\ref{f2}: (Color online) Video frames of the dusty 
plasma formation for $2a = 3.42 \times 10^{ - 4} 
\;{\mbox{cm}}$, $p_{\mathrm{Ar}} = 11\;{\mbox{Pa}}$, 
and the effective voltage between electrodes 
$U_{\mathrm{eff}} = 14.6\;{\mbox{V}}$; (a) quadrant 
camera, (b) high-resolution camera.

Fig.~\ref{f3}: (Color online) Part of the 3D dusty cloud 
scanned by the high-resolution camera. The local dust 
density obtained using the method described in 
Appendix~\ref{s9} is indicated by color.

Fig.~\ref{f4}: (Color online) Distribution of the local dust 
 density vs.\ the coordinates $x$ and $y$: (a) 3D 
view and (b) projection on the $XY$-plane.

\begin{figure}
\includegraphics[width=18cm]{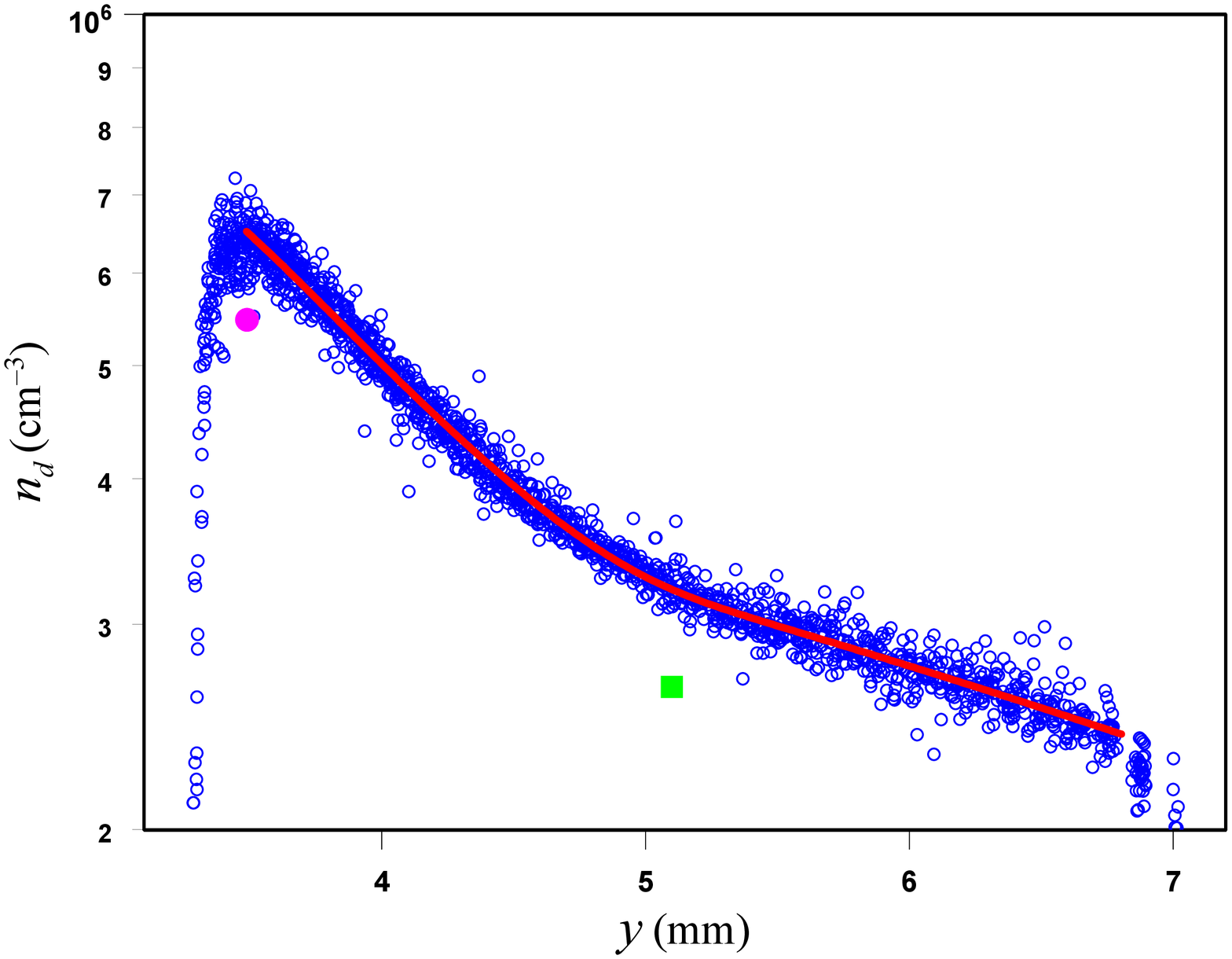}
\caption{\label{f11}}
\end{figure}

Fig.~\ref{f11}: (Color online) Dust particles number 
density ($2a = 2.55 \times 10^{ - 4} \;{\mbox{cm}}$
and $p_{\mathrm{Ar}} = 10\;{\mbox{Pa}}$) as a function 
of the distance from the discharge center. The coordinate 
axis is directed to the upper electrode. Open circles indicate 
processing the experimental data with the sampling cylinder 
radius $r_s = 0.1\;{\mbox{cm}}$
and line shows their curve fit. 
Large solid magenta circle and green square indicate theoretical
estimates for the cusp and foot particle number densities, 
Eqs.~(\ref{e18}) and (\ref{e19}), respectively.
\vskip\baselineskip

\begin{figure}
\includegraphics[width=18cm]{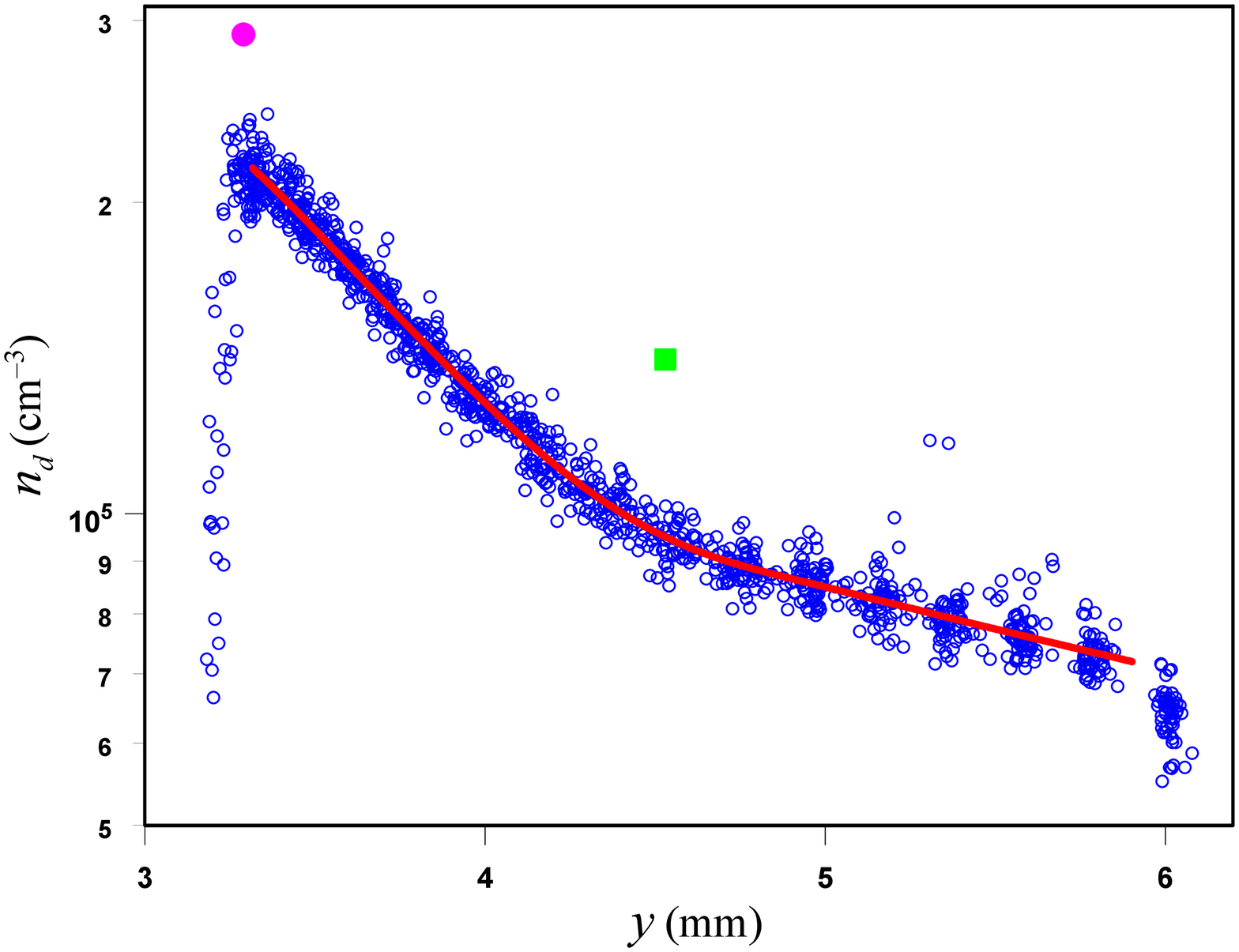}
\caption{\label{f12}}
\end{figure}

Fig.~\ref{f12}: (Color online) Same as in Fig.~\ref{f11}; 
$2a = 3.4 \times 10^{ - 4} \;{\mbox{cm}}$, 
$p_{\mathrm{Ar}} = 11\;{\mbox{Pa}}$, and $r_s = 
0.17\;{\mbox{cm}}$.
\vskip\baselineskip

\begin{figure}
\includegraphics[width=18cm]{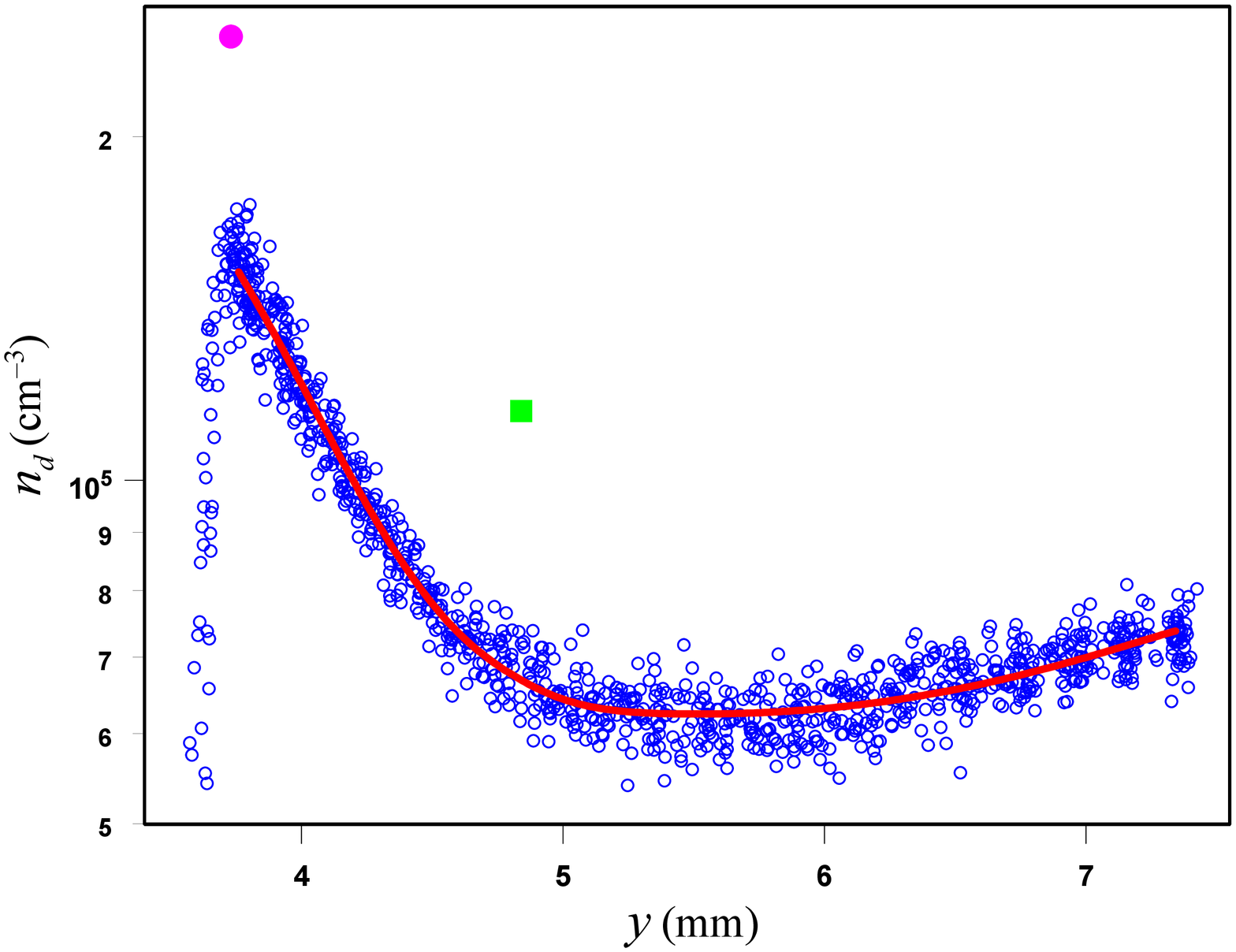}
\caption{\label{f13}}
\end{figure}

Fig.~\ref{f13}: (Color online) Same as in Fig.~\ref{f11}; 
$2a = 3.4 \times 10^{ - 4} \;{\mbox{cm}}$, 
$p_{\mathrm{Ar}} = 20.5\;{\mbox{Pa}}$, and $r_s = 
0.17\;{\mbox{cm}}$.
\vskip\baselineskip

\begin{figure}
\includegraphics[width=18cm]{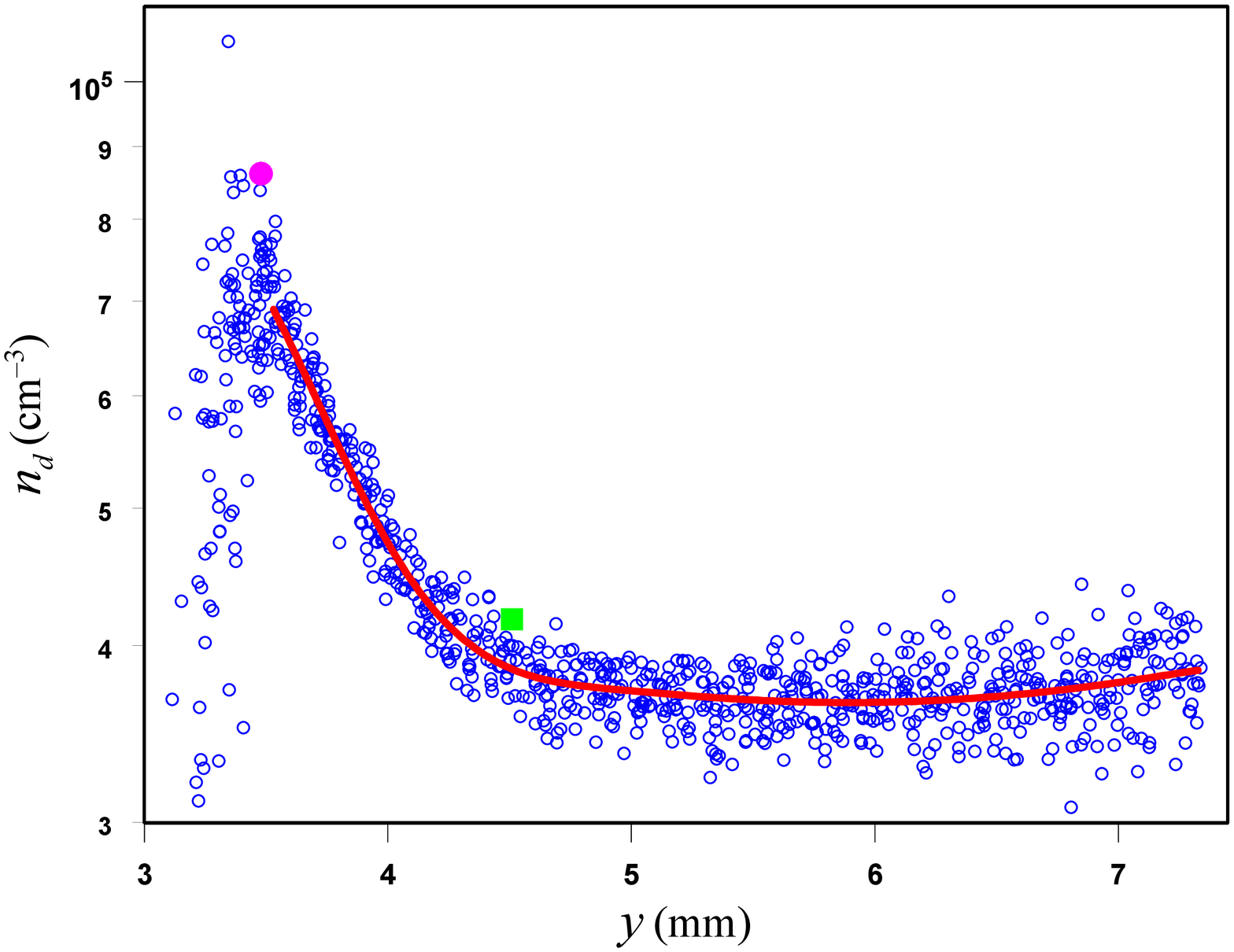}
\caption{\label{f14}}
\end{figure}

Fig.~\ref{f14}: (Color online) Same as in Fig.~\ref{f11}; 
$2a = 6.8 \times 10^{ - 4} \;{\mbox{cm}}$, 
$p_{\mathrm{Ar}} = 20.5\;{\mbox{Pa}}$, and $r_s = 
0.2\;{\mbox{cm}}$.
\vskip\baselineskip

\begin{figure}
\includegraphics[width=18cm]{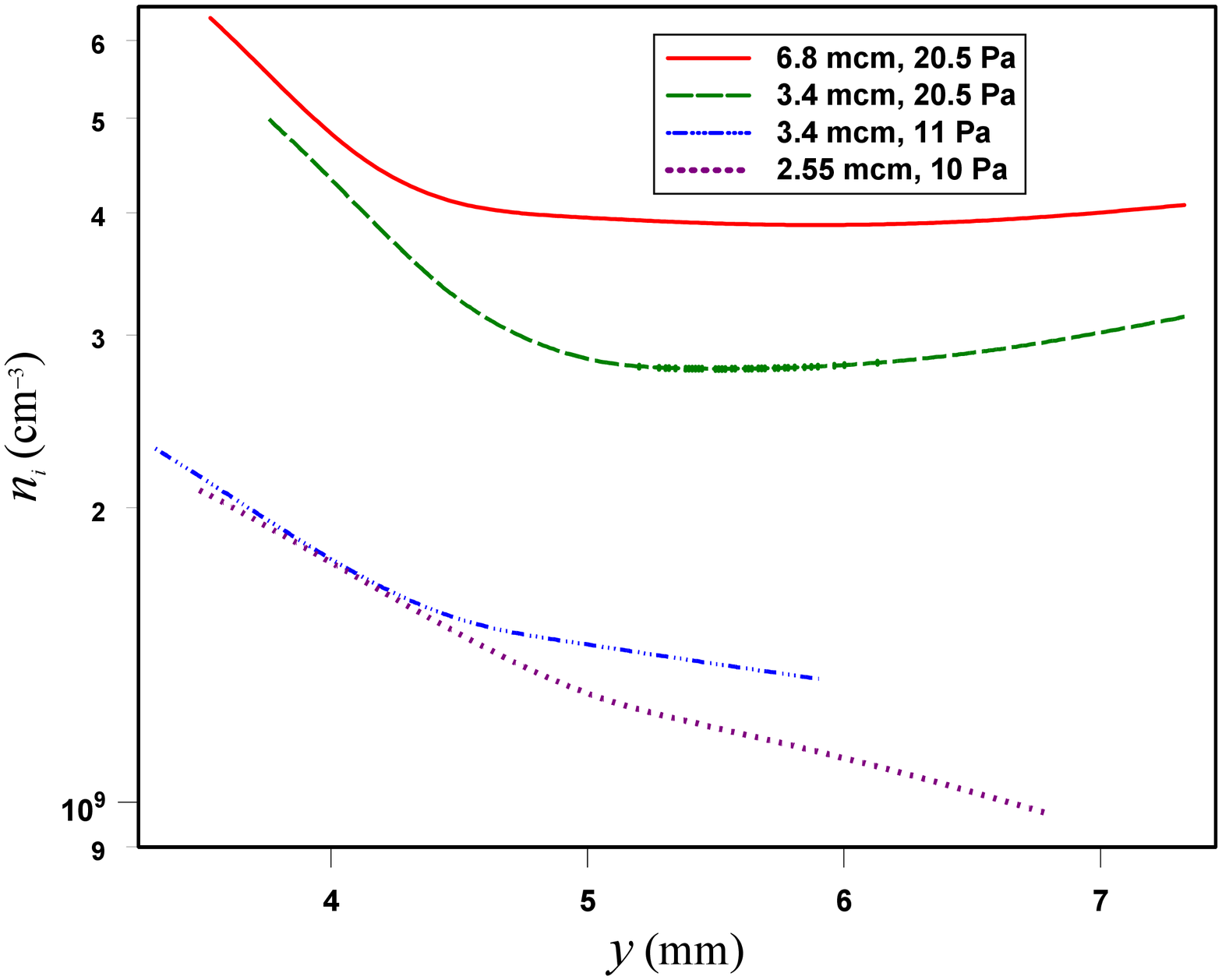}
\caption{\label{f15}}
\end{figure}

Fig.~\ref{f15}: (Color online) Ion number density for 
different particle diameters and argon pressures (see 
legend) as a function of the distance from the discharge 
center. Lines indicate the calculation using 
Eqs.~(\ref{e16}) and (\ref{e17}) (see legend).
\vskip\baselineskip

\begin{figure}
\includegraphics[width=18cm]{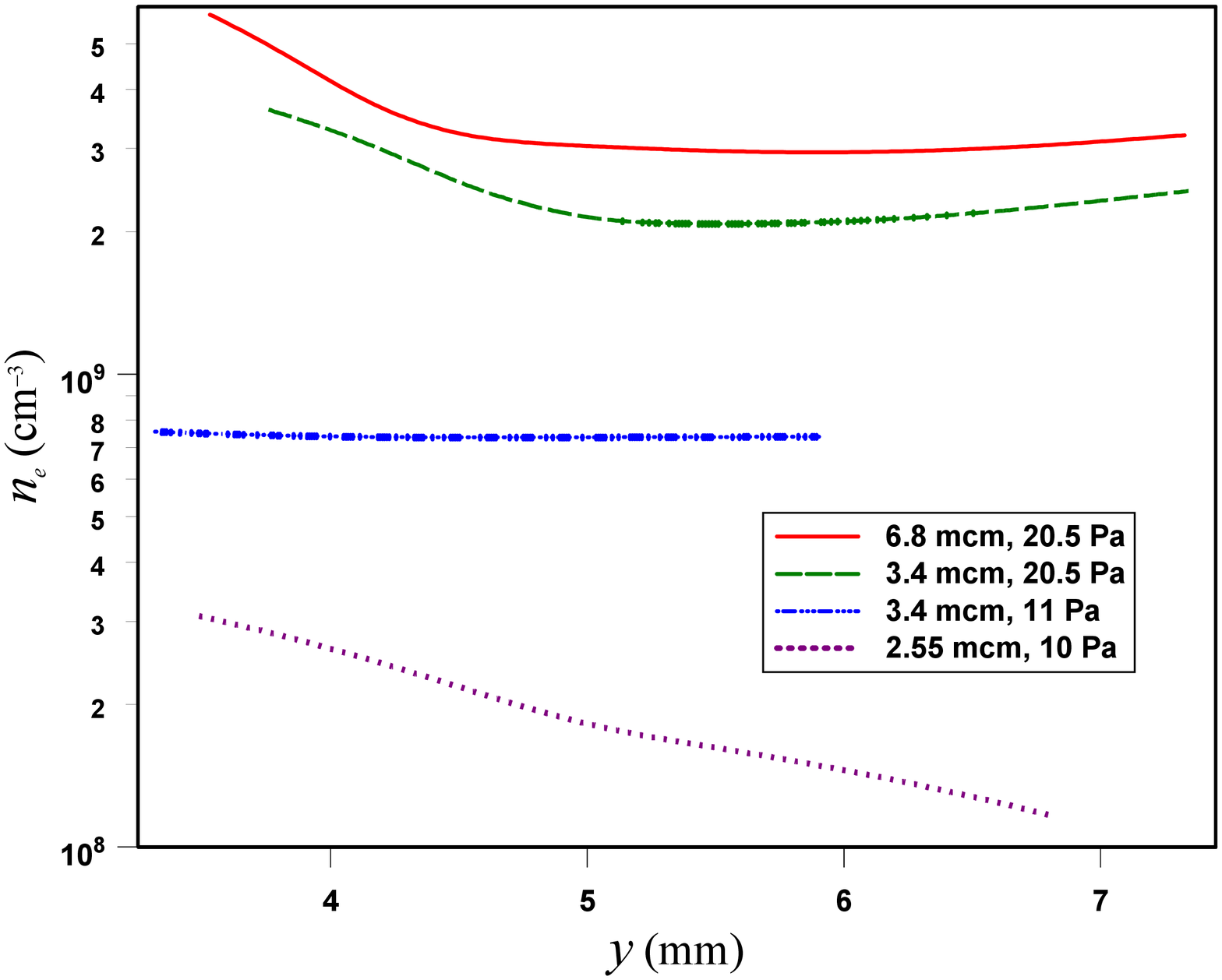}
\caption{\label{f16}}
\end{figure}

Fig.~\ref{f16}: (Color online) Electron number density for 
different particle diameters and argon pressures (see 
legend) as a function of the distance from the discharge 
center. Lines indicate the calculation using Eqs.~(\ref{e9}), 
(\ref{e15}), and (\ref{e17}) (see legend).
\vskip\baselineskip

Fig.~\ref{f_voronoi}: (Color online) View of the set of 
Voronoi cells in space. 
\vskip\baselineskip

Fig.~\ref{f_num}: (Color online) Distribution of the dust 
number density over the coordinates $x$ and $y$: (a) 3D 
view and (b) is projection on the $XY$-plane.
\vskip\baselineskip

Fig.~\ref{f_plots}: (Color online) Experimental plots of the 
dust local density in the observed region of the dust 
cloud vs.\ the coordinates $x$, $y$, and $z$.

\newpage
\vskip\baselineskip

\newpage
\vskip\baselineskip

\newpage
\vskip\baselineskip

\newpage
\vskip\baselineskip

\newpage
\vskip\baselineskip

\newpage
\vskip\baselineskip

\newpage
\vskip\baselineskip

\newpage
\vskip\baselineskip

\newpage
\vskip\baselineskip

\newpage
\vskip\baselineskip

\newpage
\vskip\baselineskip

\newpage
\vskip\baselineskip

\newpage
\vskip\baselineskip

\end{document}